\documentclass[a4paper,10pt]{article}

\usepackage[mathscr]{eucal}
\usepackage{amsmath}
\usepackage{amssymb}
\usepackage{mathtools}
\usepackage{amsthm}
\usepackage{graphicx} 
\usepackage{tempora}
\usepackage[utf8]{inputenc}
\usepackage[english]{babel}
\usepackage[T2A]{fontenc}
\usepackage[svgnames]{xcolor}
\usepackage[breaklinks,colorlinks,allcolors=DarkCyan,unicode,pdftitle={Energy transport in the Schrödinger plate}]{hyperref}
\usepackage[svgnames]{xcolor}
\usepackage[left=2cm,right=2.5cm, top=2cm,bottom=3cm]{geometry}
\usepackage{authblk}
\usepackage[sort&compress]{natbib}
\usepackage{setspace}

\theoremstyle{remark}
\newtheorem{remark}{Remark}
\newtheorem{proposition}{Proposition}

\renewcommand{\=}{\stackrel{\mbox{\scriptsize def}}{=}}
\newcommand{\I}{\text{i}}
\renewcommand{\d}{\text{d}}

\newcommand{\const}{\text{const}}

\newcommand\pd[2]{\frac{\partial#1}{\partial#2}}
\newcommand\pdd[2]{\frac{\partial^2#1}{\partial#2^2}}
\newcommand\ppdd[3]{\frac{\partial^2#1}{\partial#2\,\partial#3}}

\renewcommand\Re{\operatorname{Re}}
\renewcommand\Im{\operatorname{Im}}

\newcommand{\A}{a}

\newcommand{\phii}{{w}}

\newcommand{\tnabla}{\tilde{\bs\nabla}}
\newcommand{\bnabla}{{\bs\nabla}}
\newcommand{\tDelta}{\tilde\Delta}
\newcommand{\tDeltaa}{\bs\nabla\cdot\bs\nabla}
\newcommand{\bbn}{\bs n}
\newcommand{\bbi}{\bs i}
\newcommand{\tr}{\operatorname{tr}}
\let\bs=\boldsymbol
\let\otimess=\otimes
\renewcommand\otimes{\,}
\renewcommand\varepsilon{\mathbf E}
\renewcommand\kappa{\mathbf K}
\newcommand{\x}{x_1}
\newcommand{\y}{x_2}

\newcommand{\cddot}{\cdot\cdot}
\newcommand\tbbr{\tilde{\bs x}}
\newcommand\gammaa{(\Delta w)}
\newcommand\bbr{{\bs x}}
\newcommand\V{V}
\newcommand\LL{\mathcal L}

\title{Energy transport in the Schrödinger plate}

\author[1,2,3]{Serge N. Gavrilov}
\author[1,2,4]{Anton M. Krivtsov}
\author[1,2,3]{Ekaterina V. Shishkina}
\affil[1]{HSE University, St.~Petersburg, Russia}
\affil[2]{Yaroslav-the-Wise Novgorod State University, Veliky Novgorod, Russia}
\affil[3]{Institute for Problems in Mechanical Engineering RAS, St.~Petersburg, Russia}
\affil[4]{Peter the Great St.~Petersburg Polytechnic University (SPbPU), St.~Petersburg, Russia}

\makeatletter

\@addtoreset{equation}{section}
\@addtoreset{remark}{section}
\makeatother

\begin{document}
\selectlanguage{english}

\maketitle

\begin{abstract}
In this paper, we introduce ``the Schrödinger plate.'' This is an infinite two-dimensional
linear micro-polar elastic medium, with out-of-plane degrees of freedom, lying on a linear elastic
foundation of a special kind. Any free motion of the plate can be corresponded to a solution of the two-dimensional 
Schrödinger equation for a single particle in the external potential field $V$. The
specific dependence of the potential $V$ on the position is taken into account in the
properties of the plate elastic foundation. The governing equations of the plate are
derived as equations of the two-dimensional constraint Cosserat continuum using the direct approach. 
The plate dynamics can be described by the classical Germain--Lagrange equation for a plate, but
the strain energy is different from the one used in the classical Kirchhoff--Love
plate theory. Namely, the Schrödinger plate cannot be imagined as a thin elastic
body composed of an isotropic linear material. The main property of the
Schrödinger plate is as follows: the mechanical energy propagates in the plate exactly in
the same way as the probability density propagates according to the corresponding
Schrödinger equation.
%
%
\end{abstract}

\section{Introduction}

One hundred years ago, in 1926, Erwin Schrödinger, in a series of four papers
\cite{Schroedinger1926-AdP1,Schroedinger1926-AdP2,Schroedinger1926-AdP3,Schroedinger1926-AdP4}\footnote{The English 
translations of the Schrödinger's basic papers can be found in \cite{Schroedinger1928-collected} and
the sub-sequent editions.},
derived the equation that is considered now as the foundation
of modern quantum mechanics. In the first three papers, he argued that the new
wave mechanics should be based on the particular form of the stationary wave equation (i.e.,
the Helmholtz equation)
with variable coefficients known now as the time-independent
Schrödinger equation for a quantum particle. This equation involves an unknown parameter that has a
meaning of the particle energy.
At the beginning of the fourth paper \cite{Schroedinger1926-AdP4}, excluding
this parameter to derive the corresponding time-dependent equation, Schrödinger deduced a fourth-order
in space and second-order in time equation, which was interpreted as a
plate equation with reference to the Courant \& Hilbert book \cite{CourantHilbert1-1924}.
This equation is sometimes referred to as
Schrödinger’s real wave equation \cite{Chen1990,Chen1993,Karam2020,Makris2023}.
The factorization of the differential
operator in the plate equation obtained allowed Schrödinger to get the time-dependent
equation for a single quantum particle in the external potential field in its modern
form. Though ``the real Schrödinger wave equation'' is similar to the classical
Germain--Lagrange equation, which is used in the theory of Kirchhoff--Love plate,
it involves some additional terms.

In this paper, we try to understand to what
mechanical model this equation exactly corresponds.
We consider the two-dimensional case and 
introduce the conception of ``the Schrödinger plate.'' This is an infinite two-dimensional
linear micro-polar elastic medium, with out-of-plane degrees of freedom, lying on a linear elastic
foundation of a special kind. Any free motion of the plate can be corresponded to a solution
of the Schrödinger equation for a single quantum particle in the external field with potential $V$. The
specific dependence of the external potential $V$ on the position is taken into account in the
properties of the plate elastic foundation. The governing equations of the plate are
derived as equations of the two-dimensional constrained Cosserat continuum using the direct approach. 
The plate dynamics is described by ``the Schrödinger real wave equation''; the
additional terms appear due to the presence of the elastic foundation.
However, the plate strain energy is different from the one used in the classical Kirchhoff--Love
plate theory. We show that the Schrödinger plate cannot be imagined as a thin elastic
body composed of an isotropic linear material, i.e., as a Kirchhoff--Love
plate.

Another issue discussed in our paper is the complex nature of the
wave function. Indeed, the solution of the Schrödinger equation in its modern
form is assumed to be complex-valued, whereas the variables characterizing a 
plate oscillation are real-valued. Initially, Schrödinger assumed 
that only the real part of the wave function $\psi$ has a physical meaning; see the beginning of the original
paper \cite{Schroedinger1926-AdP4}, Schrödinger's letter to Lorenz \cite{Przibram1967}, as well as discussions in \cite{Chen1990,Chen1993,Karam2020}. 
Though a bit later he had accepted that the square of the absolute value 
$|\psi|^2=\psi\psi^\ast$
of a wave function\footnote{Here the asterisk denotes the complex conjugation},
but not a wave function itself, is a physically meaningful quantity
\cite{Schroedinger1926-AdP4,Schroedinger1928-4lectures}, 
he was not satisfied by complex wave functions; see the final remarks of 
\cite{Schroedinger1926-AdP4} and \cite{Chen1990,Chen1993}.
On the other hand, the complex nature of a
wave function is the most important issue, which makes quantum mechanics 
difficult to understand for researchers with a background mostly related to
classical mechanics. In our paper, we introduce a rather simple mechanical energetic interpretation of
the wave function. Namely, in terms of the Schrödinger plate,
its real
and imaginary parts equal, with an accuracy to the signs, the square roots of the kinetic
and potential energy densities, respectively. 
In the case of a free particle,
the similar
mechanical analogues were suggested in \cite{Chen1993} and in our previous paper 
\cite{Gavrilov2025mrc}, devoted to the Schrödinger equation in the one-dimensional case. 
{Thus, the square of the absolute value of the wave function, i.e., the quantity
proportional to the probability density according to the {Copenhagen}  (statistical) interpretation of quantum mechanics\footnote{The
Copenhagen interpretation \cite{Messiah1-1961,Stapp1997,Omnes1992} was suggested in studies of 
M.~Born, N.~Bohr, W.~Heisenberg and was never accepted \cite{Hoffmann1984} by Schrödinger.},
is equal to the total mechanical energy density for the plate.}
Hence, the main property of the
Schrödinger plate is as follows: the mechanical energy propagates in the plate exactly in
the same way as the probability density propagates according to the corresponding
Schrödinger equation. 

The relations between solutions of the Schrödinger equation and equations
describing the behaviour of plates (in 2D) and beams (in 1D) are also
discussed in studies \cite{Erofeev1992,Engstrom2023,Volovich2025,Orsingher2011,KorikovSch2021}.
The novelty of our paper is provided by
introducing the energetic interpretation of the wave function. Even for zero
potential $V$ the interpretation is not so evident, as it is discussed in
\cite{Chen1993}, and formulas suggested there are directly applicable in the 1D case of the beam only.
In the 2D case, the Schrödinger plate is not a classical Kirchhoff--Love plate. To the best of our knowledge, 
this fact was never discussed in the literature before.
We also suggest for the first time the energetic interpretation for the case of
a non-zero external potential.

The structure of the paper is as follows. 
In Sect.~\ref{sect-eq}, 
we derive basic equations of the Schrödinger plate. 
{To formulate basic equations of a linear elastic plate, we use a quite general variant 
of
the theory of plates and shells, see
\cite{Eremeyev2007,Altenbach2009,Eremeyev2006,Altenbach1988,Altenbach2004,Grekova2001,
Zhilin1976,Zhilin1982,Zhilin2006shells},
developed by the direct approach originated by study 
\cite{Ericksen1957}. According to such a variant of the plate theory, the
plate is modelled as
a two-dimensional linear elastic micro-polar Cosserat surface.
Every point of such a surface is an elementary rigid body and has six degrees
of freedom; in the three-dimensional case, the theory of the linear Cosserat continuum is developed in studies 
\cite{Cosserat1909,Palmov1964,Kafadar1971,Eringen2012}.
In Sect.~\ref{sect-eq1}, we begin with a physically clear model of 
an infinite Cosserat plane
lying on a linear elastic foundation of a special kind. The equations of motion can be uncoupled into 
the out-of-plane and in-plane systems. We consider only out-of-plane motions.
The plate elastic foundation under consideration produces both an external force 
and an external torque, which are independent.
In Sect.~\ref{sect-eq2}, we simplify the model assuming the additional
mechanical constraint, the same as used in the framework of the
Kirchhoff--Love theory ({the Kirchhoff hypothesis}). This can also be
interpreted as the transition to a model of the constraint Cosserat continuum 
 \cite{Grioli1960,Toupin1964,Mindlin1962,Aero1961,Schaefer1967,Grekova2020}
also known as the pseudo-Cosserat one. 
Since in the framework of the pseudo-Cosserat continuum we have only one external mechanical action
instead of independent force and torque,
the model of the elastic foundation, used
in the paper, becomes less physically clear.
The obtained governing equation for the
plate is the Germain--Lagrange equation for a Kirchhoff--Love plate with some
additional terms corresponding to the elastic foundation. On the other hand, this equation can be rewritten 
in the form of the real Schrödinger wave equation.
This fact is demonstrated in Sect.~\ref{S2}, where the
issues with initial conditions, which are complex-valued in the quantum
framework and real-valued in the mechanical case, are also discussed.
We show that any free motion
of the plate can be corresponded to the specific solution 
of the two-dimensional Schrödinger equation for a single particle. This can be
done in different manners; the simplest one is to correspond to the real part 
of the wave function the plate displacements in the same way as Schrödinger
did himself. Alternatively, it is possible to suggest an energetic
interpretation of the wave function.
In Sect.~\ref{sect-V=0},
transport of energy along the Schrödinger plate in the case of zero
external potential $V$ is discussed. In this case, the elastic foundation is
absent, and the mechanical energy equals
the sum of the kinetic and strain energies. The governing equation for the plate
coincides with the Germain--Lagrange equation, where only one elastic modulus
({the flexural stiffness}) is involved. At the same time, the
expression for the strain energy density involves two independent elastic
moduli. The second modulus is involved in the constitutive equation for the couple tensor only. 
For the Schrödinger plate,
we require that the energy density propagates along the plate exactly in the
same way as the probability density calculated for the wave function. This
restriction yields the specific relationship between two elastic moduli. In
Sect.~\ref{sect-KL}, we compare the expressions for the strain energies of the Schrödinger plate and 
of the classical Kirchhoff--Love plate. The latter one can be expressed in
terms of the elastic moduli for the 3D material of the plate, i.e., in terms of
the Hooke module, the Poisson ratio, and the plate thickness.
We demonstrate
that the energies exactly 
coincide, provided that the 3D material of the plate possesses a forbidden value ($\nu=1$) of the Poisson ratio.
Thus, the Schrödinger plate is not a particular case of a Kirchhoff--Love plate.
Nevertheless, in the framework of the
direct approach, the equations of the Schrödinger plate are admissible, even
though the strain energy is not positive-definite and can be zero for non-zero
strain. In
Sect.~\ref{sect-transport}, we consider the transport of energy in the more
general case of non-zero external potential. Now, the mechanical energy is the sum
of the kinetic energy, the plate strain energy, and the foundation potential
energy. In the framework of the constrained theory used for the Schrödinger
plate, the foundation potential energy loses the original meaning and can be
defined in ambiguous ways. Thus, we can introduce the modified potential energy 
such that again the total energy density propagates along the plate exactly in the
same way as the probability density calculated for the wave function.
In Conclusion (Sect.~\ref{sect-conc}) we discuss the basic results of the
paper.

The results of the paper are based on the unpublished work by A.M.~Krivtsov where 
the equivalence between the transport of the
modified energy in a beam and the probability density was established for the 1D case.
The 2D
specific results, as well as the energetic interpretation of the wave
function,
are obtained by S.N.~Gavrilov and E.V.~Shishkina.

\section{Basic equations for the Schrödinger plate}
\label{sect-eq}

\subsection{Basic equations for out-of-plane motions of a plane Cosserat surface}
\label{sect-eq1}
We mostly follow to \cite{Altenbach2009}, where the reader also can find the
extensive bibliography covering the studies where the direct approach to the theory of plates and shells
was applied.

In the framework of the approach, our plate is modelled by a two-dimensional
plane material Cosserat surface 
\begin{equation}
 S: \quad x_3=0
\end{equation}
embedded into three-dimensional space with position vector
\begin{equation}
\bbr =x_j\bs i_j. 
\end{equation}
Here, $x_j$ are Cartesian co-ordinates, 
vectors $\bs i_j$ are such that 
\begin{equation}
\bs i_j\cdot\bs i_k=\delta_{jk}, 
\end{equation}
symbol $\,\cdot\,$ defines the dot product,
$\delta_{jk}$ is the Kronecker delta. 
The position vector for points of the material surface is 
\begin{equation}
\tbbr=x_\alpha \bs i_\alpha. 
\end{equation}
Thus, 
\begin{equation}
\bs n=\bbi_3 
\end{equation}
is the normal vector for the surface. 
Here, the Einstein summation convention is assumed: Latin indices take on values $1$, $2$, or $3$; 
Greek indices take on values $1$ or $2$. 
Generally, every point of such a surface is an elementary rigid body and has six degrees
of freedom. Three of them are translational ones and correspond to the vector
of the displacement $\bs u(t,\tbbr)$, where $t$ is time. Another three are rotations
that correspond,
in the linear case, to the vector of micro-rotation~$\bs \theta(t,\tbbr)$. 
In what follows, the symbol of the tensorial product $\otimess$ will be omitted;
see, e.g., \cite{Grekova2001}. 

The linear governing equations in the differential form are \cite{Altenbach2009}:
\begin{align} 
  &\tnabla\cdot \mathbf T+\bs F=
  \rho \ddot{\bs u},
  \label{1st-E-law}
  \\
  &\tnabla\cdot \mathbf M+\mathbf T_\times+\bs L=\bs\Theta\cdot\ddot{\bs\theta}.
  \label{2nd-E-law}
\end{align} 
Here, $\tnabla=\bs i_\alpha\pd{}{x_\alpha}$ is the in-plane 2D nabla-operator, overdot is the
derivative with respect to time $t$;
$\mathbf T(t,\tbbr)$ and $\mathbf M(t,\tbbr)$ are the stress and couple tensors such that
\begin{equation}
 \bbn\cdot\mathbf T=\mathbf 0,
 \qquad
 \bbn\cdot\mathbf M=\mathbf 0;
\end{equation}
$\mathbf T_\times$ is the vectorial invariant of tensor $\mathbf T$:
\begin{equation}
  \mathbf T_\times\equiv\big(T_{jk} \bs i_j \bs i_k\big)_\times\=T_{jk} \bs i_j\times \bs i_k;
\end{equation}
$\bs F(t,\tbbr)$ and $\bs L(t,\tbbr)$ are the external force and the external torque; $\rho$ is
the mass density, symbol $\times$ denotes the cross product.
Tensor $\bs \Theta(\tbbr)$ is the inertia tensor of the elementary rigid body per unit mass, 
which is assumed to be transversely isotropic:
\begin{equation}
 \bs \Theta=\eta\, \bbn\, \bbn+ \mu \mathbf A,
\end{equation}
where
$\eta$, $\mu$ are the corresponding inertia moments,
\begin{gather}
  \mathbf A\=\tilde{\mathbf I}= \mathbf I-\bbn\otimes\bbn
\end{gather}
is the in-plane 2D identity tensor, 
$\mathbf I=\delta_{jk} \bs i_j\bs i_k$ is the 3D identity tensor.
We use notation 
\begin{gather}
\tilde {\bs z}\= \mathbf A\cdot\bs z,\\
\tilde {\mathbf Z}\=\mathbf A \cdot \mathbf Z\cdot \mathbf A
\label{Z-tilde}
\end{gather}
for an arbitrary vector $\bs z$ and a tensor of the second rank $\mathbf Z$.
Thus, the quantities with tildes are the projections of the corresponding
vectorial or tensorial quantities on the plate plane.

The constitutive equations for the stress and couple tensors are \cite{Altenbach2009}:
\begin{align}
  &\mathbf T=\pd{W}{\bs\varepsilon}
  ,
  \label{T-const-eq-pre}
  \\
  &\mathbf M=\pd{W}{\bs\kappa}
  ,
  \label{M-couple-pre}
\end{align} 
where tensors
\begin{align} 
  &\bs\varepsilon=\tnabla \bs u+ \mathbf A\times\bs\theta,
  \\
  &\bs\kappa=\tnabla \bs\theta
  \label{K-def}
\end{align} 
are 
the strain measures. Scalar quantity $W(\mathbf E,\mathbf K)$ is the strain energy density
that is assumed to be an isotropic function of its arguments.

Equations  \eqref{1st-E-law}, \eqref{2nd-E-law}, 
\eqref{T-const-eq-pre}, \eqref{M-couple-pre}
describe both out-of-plane and in-plane motions of the plate. These two types
of motions can be uncoupled. In what follows in the paper, we assume that 
the plate performs a pure out-of-plane motion. Thus, we assume that \cite{Altenbach2009}:
\begin{align}  
  &\mathbf T=
  \tilde{\bs t}\,\bbn,
  \label{t-n}
\end{align} 
where $\tilde {\bs t}$ is an in-plane vector;
\begin{align}  
  &\mathbf M=\tilde{\mathbf M}
  ,
  \\
  &\bs u=w\bbn,
  \\
  &\bs\theta=\tilde{\bs\theta},
  \\
  &\mathbf K=\tilde{\mathbf K},
  \\
  &\bs F=F\bbn,
  \\
  &\bs L=\tilde {\bs L}
  ,
  \\
  &\mathbf K=\tilde{\mathbf K}.
\end{align} 
In this case, the general form of isotropic strain energy density $W$ is
\cite{Altenbach2009}:
\begin{equation}
 2W=\alpha
 \bbn\cdot{\bs\varepsilon}^\top\cdot{\bs\varepsilon}\cdot\bbn
+\beta_1\tr{}^2\tilde{\bs\kappa}
+\beta_2\tr\tilde{\bs\kappa}^2
+\beta_3\tr\big(\tilde{\bs\kappa}\cdot\tilde{\bs\kappa}^\top\big),
\label{strain-energy}
\end{equation}
where $\alpha$, $\beta_1$, $\beta_2$, $\beta_3$ are four independent elastic moduli (the material constants), 
symbol $\tr(\cdot)$ denotes the trace of a second-rank tensor.
The constitutive equations, which follow from 
\eqref{T-const-eq-pre}, \eqref{M-couple-pre} are:
\begin{align}
  &\mathbf T
  =\alpha{\bs\varepsilon}\cdot\bbn\otimes\bbn
  =\alpha(\tnabla  w+ \tilde{\bs\theta}\times\bbn)\bbn,
  \label{T-const-eq}
  \\
  &\tilde{\mathbf M}
  =
  \beta_1\mathbf A \tr \tilde{\bs\kappa}
  +
  \beta_2 \tilde{\bs\kappa}^\top
  +
  \beta_3 \tilde{\bs\kappa}
  .
  \label{M-couple}
\end{align} 
Here, to derive the right-hand side of Eq.~\eqref{T-const-eq}, the following
relations are useful:
  \begin{gather} 
    {\bs\varepsilon}\cdot\bbn\otimes\bbn
    =\tnabla  w\bbn+ (\mathbf A \times\tilde{\bs\theta})\cdot\bbn\bbn
    =(\tnabla  w+ \tilde{\bs\theta}\times\bbn)\bbn,
  \label{enn}
  \\
  (\mathbf A \times\tilde{\bs\theta})\cdot\bbn
  =\big((\mathbf I-\bbn\otimes\bbn)\times\tilde{\bs\theta}\big)\cdot\bbn
  =(\mathbf I\times\tilde{\bs\theta})\cdot\bbn
  =-\bbn\cdot(\mathbf I\times\tilde{\bs\theta})=
  \tilde{\bs\theta}\times\bbn.
  \end{gather} 
Substituting the constitutive equations into governing equations \eqref{1st-E-law}, \eqref{2nd-E-law}
results in 
equations for displacements $w$ and micro-rotations $\tilde{\bs\theta}$:
\begin{align} 
  &\alpha(\tilde\Delta w\, \bbn+\tnabla\times\tilde{\bs\theta})+F\bbn=
  \rho \ddot{w}\bbn,
  \label{Cd1}
  \\
  &(\beta_1+\beta_2)\tnabla\tnabla\cdot\tilde{\bs\theta}+\beta_3\tilde\Delta\tilde{\bs\theta}
  +\alpha(\tnabla\times w\bbn-\tilde{\bs\theta})+\tilde {\bs L}=
  \mu
  \big(\tilde{\bs\theta}\big)^{\bs\cdot\bs\cdot},
  \label{Cd2}
\end{align} 
where $\tDelta=\tnabla\cdot\tnabla$ is the in-plane 2D Laplace operator.
To derive Eqs.~\eqref{Cd1},
  \eqref{Cd2}
we have used that
  \begin{gather} 
    \tnabla\cdot\bs\varepsilon\cdot\bbn\otimes\bbn
    =\tDelta  w\bbn+ \tnabla\cdot(\tilde{\bs\theta}\times\bbn)\bbn
    =\tDelta  w\bbn+ \bbn\cdot(\tnabla\times\tilde{\bs\theta})\bbn
    =\tDelta  w\bbn+ \tnabla\times\tilde{\bs\theta},
    \\
    (\bs\varepsilon\cdot\bbn\otimes\bbn)_\times
    =\tnabla \times w \bbn +(\tilde{\bs\theta}\times\bbn)\times\bbn 
    =\tnabla \times w \bbn -\tilde{\bs\theta}
    \label{Ennx}
  \end{gather} 
  due to Eq.~\eqref{enn}.
  The following relations are also useful:
  \begin{gather} 
    \tnabla\cdot(\mathbf A\tr\tilde{\mathbf K})=\tnabla\tnabla\cdot\tilde{\bs\theta},
    \\
    \tnabla\cdot\tilde{\mathbf K}^\top
    =\tnabla\cdot(\tnabla\tilde{\bs\theta})^\top
    =\tnabla\tnabla\cdot\tilde{\bs\theta},
    \\
    \tnabla\cdot\tilde{\mathbf K}
    =\tnabla\cdot\tnabla\tilde{\bs\theta}
    =\tDelta\tilde{\bs\theta}.
  \end{gather} 

  Now, let us introduce the plate elastic foundation. Put
\begin{gather}
 F=-k(\tbbr)w,
 \label{eforce}
 \\
 \tilde{\bs L}=-\varkappa(\tbbr) \tilde{\bs\theta}.
 \label{etorque}
\end{gather}
Here $k(\tbbr)$ and  $\varkappa(\tbbr)$ are, respectively, the translational and rotational stiffness of the plate foundation.
The external force $F$ and torque $\tilde{\bs L}$ can be associated with the external
potential energy density $\varPi$:
\begin{equation}
  2\varPi=k(\tbbr)w^2+\varkappa(\tbbr) \tilde{\bs\theta}^2.
\label{varPi}
\end{equation}
The physical meaning of the external potential $\varPi$ is given by the following
relationships:
\begin{equation}
 F=-\pd{\varPi}w,\qquad
 \tilde{\bs L}=-\pd\varPi{\tilde{\bs\theta}}.
 \label{FL-diff}
\end{equation}
The strain energy $W$ can be interpreted as the internal potential energy. Thus, in what
follows, we call the quantity 
\begin{equation}
 U=W+\varPi
\end{equation}
the potential energy density.

\subsection{Transitioning to a constrained Cosserat surface}
\label{sect-eq2}

Now, we accept the Kirchhoff kinematic hypothesis to obtain a Kirchhoff--Love--type theory.\footnote{The 
classical Kirchhoff--Love plate theory was developed using 3D
equations of linear elasticity for an isotropic material for a plane thin body \cite{love1944treatise}, whereas
we use the direct approach.}.
In the framework of the direct
approach, this hypothesis can be formally expressed as follows \cite{Zhilin2006shells,Krommer2020,Eliseev1999}:
\begin{equation}
\tilde{\bs\theta}=\tnabla\times w\bbn.
\label{constraint}
\end{equation}


\begin{remark} 
Accepting the Kirchhoff hypothesis 
\eqref{constraint} corresponds to transitioning from the model of 2D Cosserat
continuum with out-of-plane degrees of freedom to the corresponding constraint Cosserat
continuum (also known as pseudo-Cosserat continuum). 
Indeed, consider the displacement field in the neighbourhood
of the point with a position $\bs x_0$. 
Indeed, in 3D space one has
 \begin{multline} 
   \bs u (\bs x_0+\d\bbr)=
   \bs u (\bs x_0)+\d\bbr\cdot \bs\nabla\bs u=
   \bs u (\bs x_0)
   +\d\bbr\cdot (\bs\nabla\bs u)^S
   +\d\bbr\cdot (\bs\nabla\bs u)^A
   \\=
   \bs u (\bs x_0)
   +\d\bbr\cdot (\bs\nabla\bs u)^S
   -\d\bbr\cdot \mathbf I\times\bs\varphi
   =
   \bs u (\bs x_0)
   +\d\bbr\cdot (\bs\nabla\bs u)^S
   +\bs\varphi\times \d\bbr,
 \end{multline} 
 where 
 \begin{equation}
 \bs\varphi=\frac12\big((\bs\nabla\bs u)^A\big)_\times=\frac12\bs\nabla\times\bs u
 \end{equation}
 is the vector of macro-rotation;
the superscripts $S$ and $A$ denote the symmetric and antisymmetric components
of a corresponding second-rank tensor,
$\bs\nabla=\bs i_j\pd{}{x_j}$ is the in-plane 3D nabla-operator.
 For the pseudo-Cosserat continuum, it is
 accepted that the micro-rotation equals the macro-rotation 
 \cite{Grioli1960,Toupin1964,Mindlin1962,Aero1961,Schaefer1967,Grekova2020}: 
 \begin{equation}
 \bs\theta=\bs\varphi.
 \label{m=m} 
\end{equation}
 At the same time, in the two-dimensional continuum with out-of-plane degrees of freedom, one has
 \begin{multline} 
   w ({\tbbr}_0
   +\d {\tbbr})\bbn=
   w({\tbbr}_0) \bbn+\d{\tbbr}\cdot \tnabla{w}\bbn=
   w({\tbbr}_0) \bbn+\d{\tbbr}\cdot (\tnabla{w}\bbn-\bbn\tnabla w)
   =
   w({\tbbr}_0) \bbn
   -\d{\tbbr}\cdot \mathbf I\times\tilde{\bs\varphi}
   \\=
   w({\tbbr}_0) \bbn
   -\d{\tbbr}\cdot \mathbf A\times\tilde{\bs\varphi}
   =
   w({\tbbr}_0) \bbn
   +\tilde{\bs\varphi}\times \d{\tbbr},
 \end{multline} 
 where 
 \begin{equation}
   \bs\varphi\equiv\tilde{\bs\varphi}=\frac12\big(\tnabla w\bbn-\bbn\tnabla w\big)_\times=
   \tnabla\times w\bbn
 \end{equation}
 is the vector of macro-rotation.
Accepting now Eq.~\eqref{m=m} leads to Eq.~\eqref{constraint}.
\end{remark} 

To take into account Eq.~\eqref{constraint},
we put
\begin{gather}
    \tnabla \times w \bbn -\tilde{\bs\theta}
    \equiv
    (\bs\varepsilon\cdot\bbn\otimes\bbn)_\times
    \to \bs0,
    \label{qc1}
    \\
    \alpha\to\infty\label{qc2},
\end{gather}
where Eq.~\eqref{Ennx} is used.
According to Eq.~\eqref{T-const-eq},
\begin{equation}
  \mathbf T_\times=\alpha({\mathbf E}\cdot\bbn\bbn)_\times.
\end{equation}
Thus, 
$\mathbf T_\times$
cannot be found by constitutive equation \eqref{T-const-eq}
anymore. Instead, $\mathbf T_\times$ should be found by
the equation for balance of momentum~\eqref{2nd-E-law}:
\begin{equation}
  \mathbf T_\times=
  -\beta_3\tilde\Delta\tilde{\bs\theta}
-\tilde {\bs L}.
\end{equation}
where 
the relation
\begin{equation}
 \tnabla\cdot\tilde{\bs \theta}=0,
\end{equation}
which follows from Eq.~\eqref{constraint}, is taken into account as well as 
the additional simplification 
\begin{equation}
 \mathbf\Theta=\mathbf 0
\end{equation}
is taken into account.
On the other
hand, due to Eq.~\eqref{t-n}
\begin{equation}
  \mathbf T_\times=\tilde{\bs t}\times\bbn.
  \label{}
\end{equation}
Thus,
\begin{gather} 
  \tilde{\bs t}=
  (
  \beta_3\tilde\Delta\tilde{\bs\theta}
  +
  \tilde {\bs L})\times\bbn,
  \\
  \mathbf T=
  (
  \beta_3\tilde\Delta\tilde{\bs\theta}
  +
  \tilde {\bs L})\times\bbn\,\bbn,
  \label{TT-def}
\end{gather} 
\begin{multline} 
  \tnabla\cdot\mathbf T
  =
  \tnabla\cdot(\beta_3\tilde\Delta\tilde{\bs\theta}
  +
  \tilde {\bs L})\times\bbn\,\bbn
  =
  \tnabla\times(\beta_3\tilde\Delta\tilde{\bs\theta}
  +
  \tilde {\bs L})\cdot\bbn\,\bbn
  =
  \tnabla\times(\beta_3\tilde\Delta\tilde{\bs\theta}
  +
  \tilde {\bs L})
  \\=
  \tnabla\times(\beta_3\tilde\Delta\tilde{\bs\theta}
  +
  \tilde {\bs L})\cdot\bbn\,\bbn
  =
  \tnabla\times\big(\beta_3\tDelta(\tnabla\times w\bbn)
  +
\tilde {\bs L}\big)=-\beta_3\tDelta\tDelta w \bbn+\tnabla\times\tilde{\bs L}.
\end{multline} 
Thus,
Eq.~\eqref{1st-E-law} transforms into the equation for displacements, known as the 
Germain--Lagrange equation \cite{Eliseev1999,Zhilin2006shells,KorikovSch2021}
for the Kirchhoff--Love plate:
\begin{equation}
  (\beta_3\tDelta\tDelta w 
  +
  \rho \ddot{w})\bbn
  =
  \tnabla\times\tilde{\bs L}+F\bbn.
  \label{S-G}
\end{equation}
\begin{remark} 
  In the literature, traditionally, the Germain--Lagrange equation is usually derived without taking
  into account the first term in the right-hand side that corresponds to the
  external torque. The only known study for us where this term is taken into
  account is \cite{Eliseev1999}, where the Kirchhoff--Love plate theory is
  obtained by the direct approach. However, in \cite{Eliseev1999} the
  Kirchhoff--Love theory is obtained not as a limiting case of more general
  Cosserat-based theory, as we do.
\end{remark} 
The expression for the  strain energy $W$, 
according to Eqs. 
\eqref{strain-energy},
\eqref{qc1},
\eqref{qc2},
is:
\begin{equation}
 2W=
\beta_2\tr\tilde{\bs\kappa}^2
+\beta_3\tr\tilde{\bs\kappa}\cdot\tilde{\bs\kappa}^\top,
\label{strain-energy-mod}
\end{equation}
see Eq.~\eqref{strain-energy}. Here $\tilde{\mathbf K}$ is defined by Eqs.~\eqref{Z-tilde}, \eqref{K-def},
\eqref{constraint}.
The corresponding constitutive equation for the
couple tensor is 
\begin{equation}
  \tilde{\mathbf M}=
  \beta_2 \tilde{\bs\kappa}^\top
  +
  \beta_3 \tilde{\bs\kappa},
  \label{M-pseudo}
\end{equation}
 see Eq.~\eqref{M-couple}.
Now we have only two independent material constants $\beta_2,\ \beta_3$.
\begin{remark} 
One can see that though the Germain-Lagrange equation involves only one material constant
$\beta_3$, the expression for the corresponding strain energy $W$ and the
constitutive equation for $\tilde{\mathbf M}$ involve two. Note that equations for displacements \eqref{Cd1}, \eqref{Cd2} for 2D Cosserat continuum 
involve three independent material constants $\alpha$, $\beta_1+\beta_2$,
$\beta_3$ of four ones. The analogous situation takes place in the 3D
micro-polar elasticity \cite{Palmov1964,Eringen2012}.
\end{remark} 

One has
\begin{equation}
  \tnabla\times\tilde{\bs L}=-\tnabla\times(\varkappa\tnabla\times w\bbn)
 =\varkappa\tDelta w\bbn-(\tnabla\varkappa)\times(\tnabla\times w\bbn)
 =\big(\varkappa\tDelta w+(\tnabla \varkappa)\cdot\tnabla w\big)\bbn
\end{equation}
due to Eq.~\eqref{etorque}.
Thus, the right-hand side of Eq.~\eqref{S-G} is
\begin{equation}
  \tnabla\times\tilde{\bs L}+F\bbn=
\big(\varkappa\tDelta w+(\tnabla \varkappa)\cdot\tnabla w-kw\big)\bbn
\end{equation}
due to Eq.~\eqref{eforce}.
Taking into account \eqref{constraint}, Eq.~\eqref{S-G} can be rewritten as
the governing equation for the Schrödinger plate:
\begin{equation}
  \beta_3\tDelta\tDelta w 
  +
  \rho \ddot{w}
  -\big(\varkappa\tDelta w+(\tnabla \varkappa)\cdot\tnabla w\big)
  +kw=0.
  \label{S-G-inh}
\end{equation}
The last equation should be supplemented with initial conditions in the form
of
\begin{gather}
w(0,\tbbr)= w^0(\tbbr),
\qquad
\dot w(0,\tbbr)=\dot\phii^0(\tbbr),
\label{ic-2220}
\end{gather}
where $w^0(\tbbr)$, $\dot w^0(\tbbr)$ are given real-valued functions. In
this paper, we consider only an infinite plate, thus, Eq.~\eqref{S-G-inh} and initial conditions 
\eqref{ic-2220} are prescribed for all $\tbbr$. 

Expression \eqref{varPi} for the external potential $\varPi$ can now be
calculated as
\begin{equation}
  2\varPi=k(\tbbr)w^2+\varkappa(\tbbr) (\tnabla\times w\bbn)^2=
  k(\tbbr)w^2+\varkappa(\tbbr) (\tnabla w)^2.
  \label{Pi-pseudo-def}
\end{equation}
\begin{remark}  
Since in the framework of the pseudo-Cosserat continuum we have only one
external mechanical action instead of independent force and torque, i.e., the
right-hand side of Eq.~\eqref{S-G}, 
formulas like \eqref{FL-diff} are not valid anymore. Thus, the potential
$\varPi$ still can be defined for the pseudo-Cosserat continuum, but the original
physical meaning for this quantity is lost.
\label{remark-energy-meaning}
\end{remark} 

Some boundary conditions are
also required. We are only interested in localized solutions with finite
energy. Accordingly, we require 
\begin{equation}
  \iint_{-\infty}^{+\infty}\mathcal E(t,\tbbr)\,\d S<\infty
  \label{bc}
\end{equation}
for all $t\geq0$.
Here,
\begin{equation}
 \mathcal E=\mathcal K+U=\mathcal K+W+\varPi
 \label{EKU}
\end{equation}
is the total mechanical energy density,
\begin{equation}
 2\mathcal K=\rho \dot w^2
 \label{K=}
\end{equation}
is the doubled kinetic energy density.

The governing equation \eqref{S-G-inh} together with initial and boundary
conditions, \eqref{ic-2220} and \eqref{bc}, respectively, defines the
behaviour of the Schrödinger plate. However, in what follows, see Sect.~\ref{sect-V=0}, we introduce one
more restriction for such a plate, which specifies the additional relation between 
the material constants $\beta_2$ and $\beta_3$. This restriction makes
it impossible (see Sect.~\ref{sect-KL}) to consider the Schrödinger plate as a classical Kirchhoff--Love
plate on an elastic foundation.

\section{The relation of the Schrödinger plate to the Schrödinger equation}
\label{S2}
For the sake of simplicity, since only out-of-plane plate motions are under consideration, in the rest of the paper, we drop 
tildes above symbols corresponding to in-plane quantities and operators.
 
The two-dimensional time-dependent Schrödinger equation for an unknown $\varPsi$ describing behaviour of a single quantum particle with mass $m$
in the external field with
real-valued stationary potential $V(\bbr)$ has the form of:
\begin{equation}
  \I \hbar \dot{\varPsi}+\frac{\hbar^2}{2m}{\Delta\varPsi}-\V(\bbr)\varPsi=0;
\label{Schr-eq-orig}
\end{equation}
see classical textbook \cite{Messiah1-1961}. 
It can be rewritten as follows:
\begin{gather}
  \mathcal S_+\varPsi
=\I a \dot{\varPsi}+b{\Delta\varPsi}-\V(\bbr)\varPsi=0, 
\label{Schr-eq}
\\
\mathcal S_+\=\I a\frac{\partial }{\partial t}+ \LL,
\label{Sp-def}
\\
\LL= b \Delta-\V(\bbr),
\end{gather}
{where $\I$ is the imaginary unit.}
In the quantum framework,
\begin{equation}
  a=\hbar=\frac{h}{2\pi}>0, \qquad 
  b=\frac{\hbar^2}{2m}>0,
  \label{ab-def}
\end{equation}
where
$\hbar$ is the reduced Planck constant, $h$ is the Planck constant.
\begin{remark} 
Unknown $\varPsi$ in the context of the Schrödinger equation traditionally is referred to as the
quantum particle's wave function. Nevertheless, in this paper, we will call a wave function a
different quantity $\psi$ that also satisfies the Schrödinger equation. This
quantity is introduced later, in Eq.~\eqref{psi-via-chi1}.
\end{remark} 

Together with Eq.~\eqref{Schr-eq-orig} it is useful to consider the complex
conjugate equation with the operator 
\begin{equation}
\mathcal S_-\=\mathcal S_+^\ast.
\label{Sm-def}
\end{equation}
Here, the asterisk symbol denotes the complex conjugation. 
Both equations can be formulated in the following form 
\begin{gather}
\mathcal S_\pm\varPsi_\pm=
0,
\label{Schr-eq-pm}
\\
\varPsi_+\equiv \varPsi.
\end{gather}
Consider initial value problems for the Schrödinger-type equations
\eqref{Schr-eq-pm}. The initial conditions are
\begin{gather}
\varPsi_\pm(0,\bbr)=\varPsi_\pm^0(\bbr),
\label{ic}
\end{gather}
where $\varPsi_\pm^0(\bbr)$ are given complex-valued functions.
{Equations~\eqref{Schr-eq-pm} are formulated in domain $t>0$ for all $\bbr$.}
If we require that initial data for $\varPsi_\pm$ be complex conjugate: 
  \begin{equation}
  \big(\varPsi_\pm^0(\bbr)\big)^\ast=\varPsi_\mp^0(\bbr),
  \label{conjugate-conditions}
  \end{equation}
then
\begin{equation}
  \big(\varPsi_\pm(t,\bbr)\big)^\ast=\varPsi_\mp(t,\bbr),
\end{equation}
for all $t$.
\label{remark-conj}

One has
\begin{equation}
 \mathcal S_-\mathcal S_+\varPsi_++\mathcal S_+\mathcal S_-\varPsi_-=0
 \label{plate-via-S}
\end{equation}
or
\begin{gather}
\mathcal P
\phii
=0,
\label{plate-gen}
  \\
  \mathcal P
  \=\mathcal S_\pm\mathcal S_\mp
  =\mathcal S_\pm\mathcal S_\pm^\ast
  =
  \LL^2
  +
  a^2 \pdd{}{t},
\\
    \LL^2=(b\Delta-\V)^2
    =b^2\Delta^2-b\Delta\V-b\V\Delta+\V^2,
   \\
  \phii=2\Re \varPsi=\varPsi_++\varPsi_-.
\label{pchi1}
\end{gather}
Equation~\eqref{plate-gen} can be rewritten in the form 
of governing equation 
\eqref{S-G-inh} for the Schrödinger plate, i.e., in the form of 
the Germain--Lagrange equation 
\eqref{S-G}, wherein the external
force and torque are given by Eqs.~\eqref{eforce},
\eqref{etorque}, respectively.
Indeed, we have
\begin{gather}
   \begin{multlined} 
    \LL^2\varPsi_\pm=(b\Delta-\V)^2\varPsi_\pm
    =b^2\Delta^2 \varPsi_\pm -b\Delta(\V \varPsi_\pm)-b\V\Delta \varPsi_\pm+\V^2 \varPsi_\pm
    \\\qquad\qquad\qquad\qquad=
    b^2\Delta^2 \varPsi_\pm -2b(\bnabla\V)\cdot\bnabla \varPsi_\pm -2b\V\Delta \varPsi_\pm+(\V^2-b\Delta \V) \varPsi_\pm
    ,
    \label{result}
   \end{multlined} 
    \\
    \Delta(\V \varPsi_\pm)= \tDeltaa(\V \varPsi_\pm)=\bnabla\cdot\big((\bnabla\V)\varPsi_\pm+\V(\bnabla \varPsi_\pm)\big)=
    (\Delta\V)\varPsi_\pm+2(\bnabla\V)\cdot\bnabla \varPsi_\pm+\V\Delta \varPsi_\pm,
\\
  \mathcal P
  \=\mathcal S_\pm\mathcal S_\mp
  =\mathcal S_\pm\mathcal S_\pm^\ast
  =
  \A^2\frac{\partial^2 }{\partial t^2}+
    b^2\Delta^2  -2b(\bnabla\V)\cdot\bnabla  -2b\V\Delta +(\V^2-b\Delta \V). 
  \label{B-def}
  \end{gather}
  Now, taking
  \begin{equation}
  \rho=a^2,\qquad
  \beta_3=b^2,\qquad
  \varkappa=2b\V,\qquad 
  k=\V^2-b(\Delta \V)
  \label{kappa-k=}
  \end{equation}
  or
  \begin{equation}
  a=\sqrt\rho,\qquad
b=\sqrt{\beta_3},\qquad
\V=\frac\varkappa{2b},\qquad 
  k=\V^2-b(\Delta \V)
  \label{ab=}
  \end{equation}
we really rewrite Eq.~\eqref{plate-gen}
in the form of Eq.~\eqref{S-G-inh}.
\begin{remark} 
  In \cite{Volovich2025}, a similar equation is obtained by an entirely different
  approach and is interpreted in the 1D case as ``the generalized Euler-Bernoulli equation
  with potential.''
\end{remark} 
\begin{remark} 
  In terms of variables describing the Schrödinger plate, the external
  potential $V(\bbr)$ is proportional to the foundation rotational stiffness $\varkappa(\bbr)$.
\end{remark} 
Due to Eqs.~\eqref{Schr-eq}, \eqref{Sm-def}, \eqref{Schr-eq-pm}
one gets
\begin{gather}
\dot\phii(t,\bbr)=\frac{\I }{\A}\LL\big(\varPsi_+-\varPsi_+^\ast\big)=-\frac{2 }{\A}\Im\LL\varPsi(t,\bbr).
\label{pchi2}
\end{gather}
Thus, the initial conditions for $w$ are
\begin{gather}
w(0,\bbr)= w^0(\bbr)=2\Re\varPsi^0(\bbr),
\label{ic-111}
\\
\dot w(0,\bbr)=\dot\phii^0(\bbr)=-\frac{2 }{\A}\Im\LL\varPsi^0(\bbr).
\label{ic-222}
\end{gather}

\begin{remark} 
To obtain the left-hand side of the governing equation for the Schrödinger plate  \eqref{S-G-inh}, it is
enough to consider any of two terms in the left-hand side of Eq.~\eqref{plate-via-S}. 
In this way, but in the opposite direction, Erwin Schrödinger derived his
famous equation in its modern form 
\eqref{Schr-eq-orig}, see \cite{Schroedinger1926-AdP4}.
However, in the latter case,
the initial data that correspond to \eqref{ic} are generally complex, though
it should be real for a plate. Moreover, the initial values $\varPsi_\pm^0$ and $\dot\varPsi_\pm^0$ are not
independent and are related according to the Schrödinger-type equations 
\eqref{Schr-eq-pm}, though they should be independent for a plate, see
\cite{Volovich2025,Gavrilov2025mrc}. 
\end{remark} 

\begin{remark} 
  \label{remark-Psi}
As it was already discussed in Introduction, Schrödinger initially supposed
that only the real part of $\varPsi$ has a physical meaning (the plate
displacements). Although, a bit later, he discovered that $|\varPsi|^2=\varPsi\varPsi^\ast$ is 
the only meaningful quantity, he did not succeed in proposing the mechanical
analogy for this quantity \cite{Chen1993}.
\end{remark} 


To make the square of the absolute value for the solution a meaningful quantity 
in the framework of the Schrödinger plate,
introduce now the following complex valued wave function $\psi$:
\begin{equation}
\psi(t,\bbr)\=-\I\A\dot{\phii}+\LL{\phii}=\mathcal S_-w=2\LL\varPsi_+(t,\bbr),
\label{psi-via-chi1}
\end{equation}
where the last equality is due to Eqs.~\eqref{pchi1}, \eqref{pchi2}.
Since 
\begin{equation}
 \mathcal S_+\LL=\LL\mathcal S_+,
\end{equation}
function
${\psi}(t,\bbr)$ satisfies
the Schrödinger equation
\begin{equation}
\mathcal S_+\psi=0
\label{Spsi}
\end{equation}
and the initial condition, which corresponds to Eq.~\eqref{psi-via-chi1} 
considered at $t=0$, where Eqs.~\eqref{ic-111}, \eqref{ic-222} is taken into account.
Now, any motion $w(t,\bbr)$ of the Schrödinger plate, with parameters 
defined by Eq.~\eqref{kappa-k=}, can be corresponded to a
solution, i.e., the complex-valued wave function $\psi$, of the two-dimensional Schrödinger equation for a single
particle in the external potential field $V$. According to the 
{Copenhagen} interpretation,
the square $p\{\psi\}$ for the absolute value 
of $\psi$,
defined by Eq.~\eqref{E-S-earlier}, equals, with accuracy to the multiplying
constant $\lambda$ defined by Eq.~\eqref{lambda-def},
the quantum probability density for a free particle in the external field
with potential $V$. On the other hand, it is clear that 
\begin{equation}
 |\Im \psi|^2=2\mathcal K.
 \label{Im-psi}
\end{equation}
If we can take the plate parameters in
such a way that the following identity is fulfilled
\begin{equation}
|\Re \psi|^2=2U,
 \label{Re-U}
\end{equation}
then the total mechanical energy density $\mathcal E$ for the plate
equals, with accuracy to a multiplying constant, the quantum probability
density:
\begin{equation}
 p=2\lambda\mathcal E.
\label{p-E}
\end{equation}
Then, the quantum probability density $p\{\psi\}$ calculated for $\psi$ would propagate
along the plate exactly in the same way as the energy density $\mathcal E$. On
the other hand, the boundary condition \eqref{bc} transforms to the condition of square-integrability 
of the wave function that is generally accepted in the framework of the
quantum mechanics \cite{Schroedinger1926-AdP4,Messiah1-1961}:
\begin{equation}
  \iint_{-\infty}^{+\infty}\psi(t,\bbr)\psi^\ast(t,\bbr)\,\d S<\infty.
  \label{bcq}
\end{equation}

In the rest of the paper, we consider the transport of energy along the plate
and find the conditions when the energetic interpretation \eqref{Re-U}
of the real part of
the wave function is fulfilled in a certain sense.

\section{Transport of energy in the case of zero
external potential}

\label{sect-V=0}
Considering the transport of energy in the Schrödinger plate, at first, we
restrict ourselves with the case $V=0$, i.e.,
\begin{equation}
 U=W.
\end{equation}
Now, let us calculate $W$ defined by 
Eq.~\eqref{strain-energy-mod}
in Cartesian in-plane co-ordinates $x_1,\ x_2$:
\begin{proposition} 
  In Cartesian in-plane co-ordinates $x_1,\ x_2$, the terms in the right-hand
  side of expression \eqref{strain-energy-mod} for $W$ can be rewritten in the following way:
  \begin{gather}
  \tr {\mathbf K}^2=-2\pdd{w}{\x}\pdd{w}{\y}+2\left(\ppdd{w}{\x}{\y}\right)^2,
  \\
  \tr {\mathbf K}\cdot{\mathbf K}^\top=
  \left(\pdd{w}{\x}\right)^2+2\left(\ppdd{w}\x\y\right)^2+\left(\pdd{w}\y\right)^2.
  \end{gather}
  \begin{proof} 
One has 
\begin{gather}
  \tr{\mathbf K}^2
  =\tr\big( 
  (\bnabla\bnabla w\times \bbn)
  \cddot
  (\bnabla\bnabla w\times \bbn)
  \big)
  =
  (\bnabla\bnabla w\times \bbn)
  \cdot\cdot
  (\bnabla\bnabla w\times \bbn)
  ,
  \\
 \bnabla\times w \bbn=(\partial_1\bbi_1+\partial_2\bbi_2)\times w\bbn
 =-\pd{w}{x_1}\bbi_2+\pd{w}{x_2}\bbi_1,
 \end{gather}
 \begin{multline} 
   \mathbf K=\bnabla\bs\theta=\bnabla\bnabla\times w\bbn=
   (\partial_1\bbi_1+\partial_2\bbi_2)\left(-\pd{w}{x_1}\bbi_2+\pd{w}{x_2}\bbi_1\right)
 \\
 =
 -\pdd{w}{\x}\bbi_1\bbi_2
 -\ppdd{w}{\x}{\y}\bbi_2\bbi_2
 +\ppdd{w}{\x}{\y}\bbi_1\bbi_1
 +\pdd{w}{\y}\bbi_2\bbi_1,
 \label{Kii}
 \end{multline} 
 \begin{multline} 
  \tr{\mathbf K}^2
  =
  (\bnabla\bnabla w\times \bbn)
  \cdot\cdot
  (\bnabla\bnabla w\times \bbn)
=
\left(
 -\pdd{w}{\x}\bbi_1\bbi_2
 -\ppdd{w}{\x}{\y}\bbi_2\bbi_2
 +\ppdd{w}{\x}{\y}\bbi_1\bbi_1
 +\pdd{w}{\y}\bbi_2\bbi_1
 \right) 
\\\cddot
 \left(
 -\pdd{w}{\x}\bbi_1\bbi_2
 -\ppdd{w}{\x}{\y}\bbi_2\bbi_2
 +\ppdd{w}{\x}{\y}\bbi_1\bbi_1
 +\pdd{w}{\y}\bbi_2\bbi_1
 \right) 
 =
 -2\pdd{w}{\x}\pdd{w}{\y}+2\left(\ppdd{w}{\x}{\y}\right)^2 
  ,
 \label{trK2}
 \end{multline}
 \begin{multline} 
   \tr {\mathbf K}\cdot{\mathbf K}^\top
  =
  (\bnabla\bnabla w\times \bbn)
  \cdot\cdot
  (\bnabla\bnabla w\times \bbn)^\top
=
\left(
 -\pdd{w}{\x}\bbi_1\bbi_2
 -\ppdd{w}{\x}{\y}\bbi_2\bbi_2
 +\ppdd{w}{\x}{\y}\bbi_1\bbi_1
 +\pdd{w}{\y}\bbi_2\bbi_1
 \right) 
\\\cddot
 \left(
 -\pdd{w}{\x}\bbi_1\bbi_2
 -\ppdd{w}{\x}{\y}\bbi_2\bbi_2
 +\ppdd{w}{\x}{\y}\bbi_1\bbi_1
 +\pdd{w}{\y}\bbi_2\bbi_1
 \right)^\top
 =
 \left(\pdd{w}{\x}\right)^2+2\left(\ppdd{w}\x\y\right)^2+\left(\pdd{w}\y\right)^2
 .
 \label{trKKt}
\end{multline}
Here, we have used a linear (on both tensor arguments) operation \cite{Grekova2001,Lurie1990}:
\begin{equation}
 \mathbf A\cddot \mathbf B\=(\bs a_1\cdot\bs b_2)(\bs a_2\cdot\bs b_1) 
\end{equation}
if $\mathbf A=\bs a_1\bs a_2$ and $\mathbf B=\bs b_1\bs b_2$.
  \end{proof} 
\end{proposition} 

\begin{proposition} 
Strain energy $W$ defined by 
Eq.~\eqref{strain-energy-mod}
can be equivalently represented as a function of the alternative measure of strain $\mathbf K_1$:
\begin{gather}
\mathbf K_1={\mathbf K}_1=
\mathbf K_1^\top=\bnabla\bnabla w,
\\
{\mathbf K}=\mathbf K_1\times\bbn,
\end{gather}
in the following way:
\begin{gather}
 2W=c_1\tr^2\mathbf K_1+c_2\tr \mathbf K_1^2;
 \label{W-alt}
 \\
 c_1=-\beta_2,\qquad c_2=\beta_3+\beta_2;
 \\
 \beta_2=-c_1, \qquad \beta_3=c_1+c_2.
 \label{beta->c}
\end{gather}
Here, $c_1$, $c_2$ are the alternative material constants.
\end{proposition}
\begin{proof} 
According to \eqref{trKKt}, 
\begin{equation}
   \tr {\mathbf K}\cdot{\mathbf K}^\top
 =
 \left(\pdd{w}{\x}\right)^2+2\left(\ppdd{w}\x\y\right)^2+\left(\pdd{w}\y\right)^2
 =
  (\bnabla\bnabla w)
  \cdot\cdot
  (\bnabla\bnabla w).
\end{equation}
Thus, taking into account \eqref{strain-energy-mod}, \eqref{trK2}, \eqref{trKKt} one gets
 \begin{multline}
 2W=
\beta_3
 \left(\left(\pdd{w}{\x}\right)^2+2\left(\ppdd{w}\x\y\right)^2+\left(\pdd{w}\y\right)^2\right)
+\beta_2\left(
 -2\pdd{w}{\x}\pdd{w}{\y}+2\left(\ppdd{w}{\x}{\y}\right)^2 
\right)
\\=
\big((\beta_3+\beta_2)-\beta_2\big)
 \left(\left(\pdd{w}{\x}\right)^2
 +2\left(\ppdd{w}\x\y\right)^2+\left(\pdd{w}\y\right)^2\right)
+\beta_2\left(
 -2\pdd{w}{\x}\pdd{w}{\y}+2\left(\ppdd{w}{\x}{\y}\right)^2 
\right)
\\=
(\beta_3+\beta_2)
 \left(\left(\pdd{w}{\x}\right)^2+2\left(\ppdd{w}\x\y\right)^2+\left(\pdd{w}\y\right)^2\right)
-\beta_2
 \left(\left(\pdd{w}{\x}\right)^2
 +2\pdd{w}{\x}\pdd{w}{\y}
 +\left(\pdd{w}\y\right)^2
\right)
\\=
(\beta_3+\beta_2)
  (\bnabla\bnabla w)
  \cdot\cdot
  (\bnabla\bnabla w)
-\beta_2(\Delta w)^2
\equiv
c_1(\Delta w)^2
+
c_2
  (\bnabla\bnabla w)
  \cdot\cdot
  (\bnabla\bnabla w).
\label{strain-energy-mod1}
\end{multline}
\end{proof} 
\begin{remark} 
  It may be shown that constitutive equation 
  \eqref{M-pseudo} can be rewritten in terms of the measure of strain $\mathbf K_1$
  the and elastic moduli $c_1$ and $c_2$ in the following way
  \cite{Eliseev1999}:
\label{remark-M-K1}
\begin{equation}
 \mathbf M=c_1 (\tr \mathbf K_1) \mathbf A\times \mathbf n +c_2\mathbf K_1\times
 \mathbf n.
 \label{M-K1}
\end{equation}
\end{remark} 
Alternative representation \eqref{W-alt} is more useful when considering
transport of energy in the Kirchhoff--Love and Schrödinger plates. It is used
in \cite{Eliseev1999}, where the Kirchhoff--Love plate is considered by the
direct approach. The choice of the alternative constants also allows us to
obtain the formulas describing the energy transport in the form analogous to one 
derived in \cite{Gavrilov2025mrc} for the 1D case.

Provided that $V=0$, operator
$\mathcal P$ defined by Eq.~\eqref{B-def} can be rewritten as follows:
  \begin{gather}
    \mathcal P\big|_{V=0}
  =
  \rho\frac{\partial^2 }{\partial t^2}+
  (c_1+c_2)\Delta^2.  
  \label{P0-def}
  \end{gather}
Here, Eqs.~\eqref{ab=}, \eqref{beta->c} are taken into account.

Assuming that $\phii$ is a real valued function,
one can obtain the equation for the balance of energy in the plate by means of
multiplying Eq.~\eqref{plate-gen} by $\dot w$. After some transformations, we
get
\begin{equation}
\dot{\mathcal E}=-\bnabla\cdot \bs Q, 
\label{e-balance}
\end{equation}
where 
\begin{gather}
  2 \mathcal E
  =2\mathcal K+2W
  =
  a^2
  \dot w^2
  +
  c_1
  (\Delta w)
  ^2
  +
  c_2
  \bnabla\bnabla w\cddot\bnabla\bnabla w
  ,
  \label{E-BE-V}
  \\
  \bs
  Q
  =\bs Q_1
  +\bs Q_2,
  \\
  \bs Q_1=
  c_1\big(\dot w(\bnabla\Delta w)
  -\gammaa(\bnabla\dot w)\big)
  ,
  \\
  \bs Q_2=c_2
  \big(\dot w(\bnabla\Delta w)-(\bnabla\bnabla w)\cdot(\bnabla\dot w)\big).
\end{gather}
Here, 
$\mathcal E$
is the mechanical energy density,
$\bs Q$ is the
mechanical energy flux,
which corresponds to the solution $\phii$ of the Germain-Lagrange equation.
Indeed,
\begin{gather} 
\begin{multlined} 
\dot{\mathcal E}=
a^2\dot w\ddot w+
c_1\big(\Delta w\big)\big( \Delta \dot w\big)+c_2
\bnabla\bnabla w\cddot\bnabla\bnabla \dot w 
\\\qquad=
c_1\big((\Delta w)(\Delta \dot w)-
\big(
\Delta^2w\big)\dot w
\big)
+c_2
\big(\bnabla\bnabla w\cddot\bnabla\bnabla \dot w-(\Delta\Delta w)\dot w\big) 
=-\bnabla\cdot(\bs Q_1+
\bs Q_2),
\label{dot-E-V0}
\end{multlined} 
\\
c_1\big(\gammaa(\Delta \dot w)-(\Delta\Delta w)\dot w\big)
=c_1\bnabla\cdot\big(\gammaa(\bnabla\dot w)-\dot w(\bnabla\Delta w)\big)
=-\bnabla\cdot \bs Q_1,
\label{nablaQ1}
\\
c_2\big(\bnabla\bnabla w\cddot\bnabla\bnabla \dot w
-(\Delta\Delta w)\dot w\big)
= c_2\bnabla\cdot\big((\bnabla\bnabla w)\cdot(\bnabla\dot w)-\dot w(\bnabla\Delta w)\big)
=-\bnabla\cdot \bs Q_2.
\end{gather} 
Here, when calculating the right-hand side of 
Eq.~\eqref{dot-E-V0},
Eq.~\eqref{plate-gen} is taken into account.

According to  Eqs.~\eqref{e-balance}, \eqref{bc}
\begin{equation}
  \frac\d{\d t} \left(\iint_{-\infty}^{+\infty}\mathcal E(t,\bbr)\,\d S\right)=0
  \label{conserves}
\end{equation}
provided that $\bs Q\to 0$ as $\bbr\to\infty$, i.e., the global mechanical energy of the plate is conserved
due to the Ostrogradsky--Gauss theorem.

Let us calculate the quantum 
probability density $p\{\psi\}$ defined by Eq.
\eqref{E-S-earlier},
where  $\psi$ is  
defined by Eq.~\eqref{psi-via-chi1}.
According to Eqs.~\eqref{ab=}, \eqref{beta->c}
we have
\begin{gather}
  p
  =
  \lambda\big(\rho \dot w^2+(c_1+c_2)(\Delta w)^2\big)
  .
  \label{E=}
\end{gather}
{Comparing Eq.~\eqref{E-BE-V} with Eq.~\eqref{E=},
one can see that Eq.~\eqref{p-E} is fulfilled 
if and only if the strain energy for the Schrödinger plate is such that}
\begin{equation}
 c_2=0,
 \label{c2=0}
\end{equation}
i.e.,
\begin{gather}
 2W
 =
 c_1\big(\Delta w\big)^2
 =
 b^2\big(\Delta w\big)^2
 \label{strain-energy-equiv}
\end{gather}
due to Eqs.~\eqref{ab=}, \eqref{beta->c},
and
\begin{equation}
  \mathbf M=
  b^2(\Delta w)\, \mathbf A\times \bs n
  \label{M-S}
\end{equation}
due to Eq.~\eqref{M-K1}.
In the last case, the mechanical energy for the Schrödinger plate with density 
$\mathcal E=\mathcal K+U$
and the corresponding probability density $p$ propagate in the same way. In terms of $p$,
equation for the balance of energy 
\eqref{e-balance} can be rewritten now {in the form} of 
Eqs.~\eqref{rho-balance-earlier}, \eqref{q-def}.

\begin{remark} 
  It is easy to show that fluxes $\bs Q$ and $\bs q$ are also related by the
  formula analogous to Eq.~\eqref{p-E}:
\begin{equation}
 \bs q=2\lambda \bs Q.
\end{equation}
\end{remark} 
%

\section{Comparison with the Kirchhoff-Love plate}
\label{sect-KL}
In the framework of the classical Kirchhoff--Love plate theory, which is developed using 3D
equations of linear elasticity for an isotropic material for a plane thin body, the equation of a
plate motion is \cite{love1944treatise}:
\begin{equation}
 \mathcal D\Delta\Delta w+\rho \ddot w=F,
\end{equation}
where
\begin{equation}
  \mathcal D=\frac1{12}\frac{Eh^3}{1-\nu^2}
\end{equation}
is the flexural stiffness.
The corresponding expression for the strain energy is \cite{love1944treatise,Courant1989-1,Ventsel2001}:
\begin{equation}
 2\mathcal W=\mathcal D\left((\Delta w)^2+(1-\nu)
\left(-2\pdd{w}{\x}\pdd{w}{\y}+2\left(\ppdd{w}{\x}{\y}\right)^2\right)
 \right),
\end{equation}
where
 $\nu$ is the Poisson ratio for the 3D material, $E$ is the Young
 modulus for the 3D material, $h$ is the plate thickness.
 One has:
 \begin{equation}
 2\mathcal W
 =
 \mathcal D\left(
   \underbrace{\tr \big({\mathbf K}\cdot{\mathbf K}^\top\big)-
   \tr{\mathbf K}^2}_{(\Delta w)^2}
  +(1-\nu)\tr{\mathbf K}^2
 \right)
 =
 \mathcal D\left(
   \tr \big({\mathbf K}\cdot{\mathbf K}^\top\big)-
  \nu\tr{\mathbf K}^2
 \right).
\label{strain-energy-Love}
 \end{equation}
Taking $W=\mathcal W$, 
and comparing Eqs.~\eqref{strain-energy-mod}
\& \eqref{strain-energy-Love} one gets:
\begin{equation}
 \mathcal D=
 \beta_3=
 c_2+c_1
 ,\qquad 
 \nu=-\frac{\beta_2}{\mathcal D}=\frac{c_1}{c_1+c_2}. 
 \label{beta32}
\end{equation}
For the Schrödinger plate Eq.~\eqref{c2=0} is fulfilled, and thus 
%
\begin{equation}
\nu=1.
\label{nu1}
\end{equation}
Recall that in linear isotropic elasticity the value of the Poisson ratio
should satisfy the inequality
\begin{equation}
 -1<\nu<1/2,
\end{equation}
to have the strain energy of 3D elastic material be positive-definite. 
Thus, the value of the Poisson ratio 
\eqref{nu1} is unacceptable \cite{Lurie2005}. 
{Nevertheless, the strain energy defined by Eq.~\eqref{strain-energy-equiv}
is acceptable in the framework of the direct approach.}
Thus, the Schrödinger plate is not a particular case of a Kirchhoff--Love plate.

\begin{remark} 
It is easy to show that the strain energy defined by 
Eq.~\eqref{strain-energy-mod1} is positive-definite if and only if
the following conditions are fulfilled:
\begin{equation}
 \beta_3>0\qquad\text{and}\qquad\beta_3+\beta_2>0\qquad\text{and}\qquad\beta_3-\beta_2>0.
 \label{positive}
\end{equation}
  \begin{proof} 
   According to  Eq.~\eqref{strain-energy-mod1},
   \begin{gather}
    2W=\mathcal X^\top\mathcal M \mathcal X,
    \\
    \mathcal X=
    \begin{pmatrix} 
      \ppdd{w}\x\y
      \\
      \pdd w\x
      \\
      \pdd w\y
    \end{pmatrix},
    \qquad
    \mathcal M=
    \begin{pmatrix} 
      2(\beta_3+\beta_2) & 0 &0
      \\
      0&\beta_3 & -\beta_2
      \\
      0 & -\beta_2 & \beta_3
    \end{pmatrix}.
   \end{gather}
   Applying the Sylvester criterion, see, e.g., \cite{Akivis1972}, one gets that the matrix $\mathcal M$ is
    positive-definite if and only if inequalities  \eqref{positive} are
    satisfied.
  \end{proof} 
\end{remark}

\begin{remark} 
  \label{remak-nonn}
  The strain energy defined by Eq.~\eqref{strain-energy-equiv} is not
  positive-definite but non-negative-definite. It can be zero for $\bnabla\bnabla w\neq\mathbf0$, since $\beta_3$ and $\beta_2$ defined by
  Eqs.~\eqref{beta32}, \eqref{nu1} break the second inequality~\eqref{positive}: $\beta_3+\beta_2=0$.
  An example of $w\neq0$ such that $\bnabla\bnabla w\neq\mathbf0$ but $W$ defined by 
 Eq.~\eqref{strain-energy-equiv}
  equals zero is
  \begin{equation}
    w=x_1x_2.
  \end{equation}

\end{remark} 
\section{Transport of energy in the case of non-zero
external potential}
\label{sect-transport}
In the case $V\neq0$, we deal with the Schrödinger plate on the elastic
foundation and expect that the total mechanical energy with density
$\mathcal E=\mathcal K+W+\varPi$
is conserved:
\begin{equation}
  \frac\d{\d t} \left(\iint_{-\infty}^{+\infty}\mathcal E(t,\bbr)\,\d S\right)=0.
  \label{conserves-V}
\end{equation}
The strain energy density $W$ is defined by 
Eq.~\eqref{strain-energy-equiv}.
The potential energy associated with the elastic
foundation $\varPi$ is defined by the following relation
\begin{equation}
 2\varPi=(V^2-b\Delta V)w^2+2bV(\bnabla w)^2
 \label{P=}
\end{equation}
due to Eqs.~\eqref{Pi-pseudo-def}, \eqref{kappa-k=}. Thus,
\begin{gather}
 2\mathcal E=
  {\A^2} \dot w ^2
  +
 b^2\big(\Delta w\big)^2
 + 
 (V^2-b\Delta V)w^2+2bV(\bnabla w)^2.
 \label{E=K+W+P-mod}
\end{gather}

One can obtain the equation for the balance of energy $\mathcal E$ by means of
multiplying Eq.~\eqref{plate-gen} by $\dot w$. After some transformations, we
get
\begin{equation}
\dot{\mathcal E}=-\bnabla\cdot \bs Q, 
\label{e-balance-V}
\end{equation}
where 
\begin{gather}
  \bs
  Q\{\phii\}=\bs Q_1+\bs Q_V
  ,
  \\
\bs Q_V=
 -2b\V (\bnabla w) \dot w
.
\end{gather}
Here, $\bs Q$ is the flux for energy with density $\mathcal E$.
\begin{proof} 
\begin{multline} 
\dot{\mathcal E}=
a^2\dot w\ddot w+
b^2\big(\Delta w\big) \big(\Delta\dot w\big)+
 (V^2-b\Delta V)w
 \dot w+2bV(\bnabla w)(\bnabla \dot w)
\\=
 - \bnabla \cdot\bs Q_1
 +(V^2-b\Delta V)w\dot w+2bV(\bnabla w)\cdot\bnabla \dot w
 +\big(2b (\bnabla V) \cdot  (\bnabla w)\dot w
 +2b V(\Delta w)
 -(V^2-b\Delta V)w
 \big)\dot w
 \\
 =
 - \bnabla \cdot\bs Q_1
 +2b(\bnabla V)\cdot(\bnabla w) \dot w
 +2bV(\Delta w )\dot w
 +2bV(\bnabla w)\cdot\bnabla \dot w
 =
 - \bnabla \cdot(\bs Q_1
 +\bs Q_V)
 .
\end{multline} 
Here, Eqs.~\eqref{dot-E-V0}, 
\eqref{nablaQ1} were used.
\end{proof} 
Thus, provided that $\bs Q\to0$ as $\bbr\to\infty$, the global mechanical
energy with density $\mathcal E$ is conserved
due to the Ostrogradsky--Gauss theorem, i.e., Eq.~\eqref{conserves-V} is fulfilled.

\begin{remark} 
  Note that transport of energy in a plate with the elastic foundation of a similar
  structure was considered in study 
  \cite{Erofeev2023}.
\end{remark} 
Let us now calculate the quantum 
probability density $p\{\psi\}$ defined by Eq.~\eqref{E-S-earlier},
where  $\psi$ is given by Eq.~\eqref{psi-via-chi1}.
According to Eqs.~\eqref{ab=}, \eqref{beta->c}, \eqref{c2=0},
we have
\begin{gather}
  p
  =
  \lambda\Big(a^2 \dot w^2+
  \big((b\Delta-\V)w\big)
  ^2
\Big)
  .
  \label{E=V}
\end{gather}
Comparing Eq.~\eqref{E=V} with Eq.~\eqref{E=K+W+P-mod} wherein the terms in the right-hand sides are
defined by Eqs.~\eqref{K=}, \eqref{strain-energy-equiv}, \eqref{P=}, respectively, one can see that 
\begin{equation}
 p\neq2\lambda\mathcal E
 \label{p-E-V0}
\end{equation}
as we have according to Eq.~\eqref{p-E} for $V=0$\footnote{The proof of this fact is clear from the following text.}. 
Since for the Schrödinger
plate the original physical
meaning for the external potential energy is lost (see Remark~\ref{remark-energy-meaning}), and the energy 
now is only a quantity, which is globally conserved (see Eq.~\eqref{conserves-V}) we can introduce 
a modified energy with a density 
\begin{equation}
 \mathcal E^M=\mathcal E-\bnabla\cdot \bs d=\mathcal K+W+\varPi^M,
 \label{EM-def}
\end{equation}
where $\bs d$ is such that 
\begin{equation}
  \bs d\to0, \quad \dot{\bs d}\to 0 \quad\text{as}\quad {\bs \bbr}\to\infty 
  \label{d-to-0}
\end{equation}
and
\begin{equation}
 p=2\lambda\mathcal E^M, \qquad \lambda=\const.
\end{equation}
Here,
\begin{equation}
 \varPi^M=\varPi-\bnabla\cdot \bs d
 \label{PiM-def}
\end{equation}
is the modified potential energy of the foundation.
The global mechanical energy with density $\mathcal E^M$ for the Schrödinger plate is conserved:
\begin{equation}
  \frac\d{\d t} \left(\iint_{-\infty}^{+\infty}\mathcal E^M(t,\bbr)\,\d S\right)=0, 
\end{equation}
and the probability density $p$ propagates in the same way as
$\mathcal E^M$ likewise it has been obtained in Sect.~\ref{sect-V=0} for the case $V=0$. Moreover,
global energies corresponding to densities $\mathcal E$ and $\mathcal E^M$
are equal to each other:
\begin{equation}
  \iint_{-\infty}^{+\infty}\mathcal E(t,\bbr)\,\d S=
  \iint_{-\infty}^{+\infty}\mathcal E^M(t,\bbr)\,\d S
  ,
\end{equation}
since 
\begin{equation}
  \iint_{-\infty}^{+\infty}\bnabla \cdot \bs d\,\d S=0
\end{equation}
due to \eqref{d-to-0}.
\begin{proposition} 
  Let 
  \begin{equation}
    \bs d=-\frac b2\bnabla(Vw^2)+bV\bnabla(w^2).
  \end{equation}
  Then $\mathcal E^M$ defined by Eq.~\eqref{EM-def} is such that
  \begin{equation}
  2 \mathcal E^M=
  {\A^2}
  \dot w
  ^2
  +
  \big((b\Delta-\V)w\big)
  ^2.
  \label{E-BE-VM}
  \end{equation}
\end{proposition} 
\begin{proof} 
  \begin{gather} 
    \mathcal E^M-\mathcal E=-bVw\Delta w+\frac b2(\Delta V)w^2-bV(\bnabla w)^2
    =-\bnabla \cdot \bs d.
  \end{gather} 
\end{proof} 
\begin{proposition} 
The equation for balance of energy with density $\mathcal E^M$ can be
written as follows:
\begin{equation}
\dot{\mathcal E}^M=-\bnabla\cdot \bs Q^M, 
\label{e-balance-VM}
\end{equation}
where 
\begin{gather}
  \bs
  Q^M=\bs Q_1+\bs Q_V^M
  ,
 \\
  \bs Q^M_V
    =
 b\big(\V w\bnabla \dot w
 -V(\bnabla w) \dot w
 -(\bnabla\V) w\dot w
\big)
.
\end{gather}
Here, $\bs Q^M$ is the  flux for energy with density $\mathcal E^M$.
\end{proposition} 
\begin{proof} 
\begin{gather} 
\begin{multlined} 
\dot{\mathcal E}^M=
a^2\dot w\ddot w+
\big((b\Delta-\V)w\big) \big((b\Delta -\V)\dot w\big)
\\\qquad\qquad\qquad\qquad
=
\big((b\Delta -V)w\big)\big((b\Delta -V)\dot w\big)-
\big(
(b\Delta-\V)^2w\big)\dot w
=-\bnabla\cdot(\bs Q_1+\bs Q_V^M
),
\end{multlined} 
\\
\begin{multlined} 
\big((b\Delta -V)w\big)\big((b\Delta -V)\dot w\big)-
\big(
(b\Delta-\V)^2w\big)\dot w
\\
\qquad\qquad\qquad\qquad
=-\bnabla\cdot \bs Q_1
+b\big(\V(\Delta w)\dot w-\V w\Delta \dot w+2(\bnabla \V)\cdot(\bnabla w) \dot
w+(\Delta \V) w\dot w\big),
\end{multlined} 
\\
\begin{multlined} 
b\big(\V(\Delta w)\dot w-\V w\Delta \dot w+2(\bnabla \V)\cdot(\bnabla w) \dot w+(\Delta \V) w\dot w\big)
\\
\qquad\qquad\qquad\qquad
=-b\bnabla\cdot\big(\V w\bnabla \dot w
 -V(\bnabla w) \dot w
 -(\bnabla\V) w\dot w
 \big)
=-\bnabla\cdot \bs Q_V^M
.
\end{multlined} 
\end{gather} 
Here, Eqs.~\eqref{dot-E-V0}, 
\eqref{nablaQ1} were used.
\end{proof} 

Thus, the modified mechanical energy for the Schrödinger plate with density 
$\mathcal E^M=\mathcal K+W+U^M$
and the corresponding probability density $p$ propagate in the same way. In terms of $p$,
equation for the balance of energy 
\eqref{e-balance} can be rewritten now {in the form} of 
Eqs.~\eqref{rho-balance-earlier}, \eqref{q-def}.

\begin{remark} 
  It is easy to show that fluxes $\bs Q^M$ and $\bs q$ are also related by the
  formula analogous to Eq.~\eqref{p-E}:
\begin{equation}
 \bs q=2\lambda \bs Q^M.
\end{equation}
\end{remark}

\section{Conclusion}
\label{sect-conc}
In this paper, we have introduced 
``the Schrödinger plate'' and have shown that any motion of such
a plate can be corresponded to a solution of the Schrödinger equation 
\eqref{Schr-eq-orig}
for a quantum particle subjected to the external potential.
The
specific dependence of the external potential on the position is taken into account in the
properties of the plate elastic foundation. Namely, the external potential is related to both 
translational and rotational stiffnesses of the
plate foundation and is proportional to the rotational one; see Eq.~\eqref{kappa-k=}.

The
correspondence that we discuss can be established in various ways. The first way, given by Eq.~\eqref{pchi1},
is to relate the plate displacements with the real part of
a solution $\varPsi$  of the Schrödinger equation. Essentially, this was the way Schrödinger
himself derived his famous equation. In such a way, the imaginary  part
$\Im \varPsi$ also got the mechanical interpretation, which can be given by
Eq.~\eqref{pchi2}. However, the only meaningful quantity $|\varPsi|^2=\varPsi\varPsi^\ast$ has 
not got any clear mechanical interpretation; see Remark~\ref{remark-Psi}. 
There is the alternative energetic interpretation. Any
motion of the Schrödinger plate can be corresponded to the wave function $\psi$ defined by 
Eq.~\eqref{psi-via-chi1}, which also satisfies the Schrödinger equation. The
imaginary part clearly satisfies Eq.~\eqref{Im-psi}, and we planned to check
if the real part satisfies Eq.~\eqref{Re-U}. Hence, in such a way, we would
suggest a mechanical system where the total mechanical energy
propagates exactly in the same way as the quantum probability density 
\eqref{E-S-earlier}.  

Our plan was successfully realized in the case of the
Schrödinger equation for a free quantum particle with zero external potential
$V=0$; see Sect.~\ref{sect-V=0}.
We have proved that it is possible to choose the plate parameters in such a
way that Eq.~\eqref{Re-U} is fulfilled. It is interesting that to do this, we
need to choose the strain energy for the Schrödinger plate in the form, which
is not admissible for the classical Kirchhoff--Love plate; see Sect.~\ref{sect-KL}.
\footnote{Although the
motions of the  Schrödinger plate and Kirchhoff--Love plate are governed by
the same Germain--Lagrange equation in the case $V=0$.} Moreover, the strain
energy for  the Schrödinger plate is not positive-definite (but
non-negative-definite; see Remark~\ref{remak-nonn}).

In the case of a non-zero external potential $V$ (see Sect.~\ref{sect-transport})
we did not succeed in choosing the
parameters in such a way that Eq.~\eqref{Re-U} is fulfilled. However, we
achieve our goal by introducing the modified potential energy $\varPi^M$ of the
foundation \eqref{PiM-def}. This energy ``is not worse'' than the original energy $\varPi$. Indeed, 
the corresponding total mechanical energy $\mathcal E^M$
\eqref{EM-def} is also globally conserved, whereas the original physical
meaning for the external potential energy is lost; see Remark~\ref{remark-energy-meaning}. 
Thus, we have successfully suggested the mechanical system where the energy
propagates in the same way as the probability density, and this is the
main result of the paper. Note that the paper can be formally rewritten in the 1D case, 
and in such a way it provides the mechanical interpretation for the
1D Schrödinger equation with non-zero external potential, extending the result
of our previous paper \cite{Gavrilov2025mrc}, where the case $V=0$ was
considered. It would be interesting to check if the analogous mechanical interpretation can be obtained
in the 3D case, where we have the pseudo-Cosserat continuum instead of the Schrödinger plate. 

We would like to emphasize that both theories, i.e., 
the classic quantum mechanics where the Schrödinger equation 
is applicable and the theory of the Schrödinger plate, are some approximations of more
general theories where the perturbations propagate at a limiting speed, i.e.,
of the relativistic quantum mechanics and non-classical Cosserat-based plate theories,
respectively.

Finally, we hope to achieve deeper analogies
between the wave propagation in a plate and corresponding solutions of the
time-dependent Schrödinger equation in the framework of the energy dynamics.
The energy dynamics was recently suggested 
\cite{Krivtsov2022,Kuzkin2023,Baimova2023}
as a framework that allows us to
introduce wave–particle duality into  classical mechanics in a rational
way.

\section*{Author contribution statement}
The results of the paper are based on the unpublished work by A.M.~Krivtsov where 
the equivalence between the transport of the
modified energy in a beam and the probability density was established for the 1D case.
The 2D
specific results, as well as the energetic interpretation of the wave
function,
are obtained by S.N.~Gavrilov and E.V.~Shishkina. The manuscript draft is written by S.N.~Gavrilov.
A.M.~Krivtsov and E.V.~Shishkina have provided the review and editing of the
manuscript.

\section*{Acknowledgement}
{The authors are grateful to 
A.A.~Sokolov who attracted our attention to this problem, and to
H.~Altenbach,
V.A.~Eremeyev,
E.F.~Grekova, 
E.A.~Ivanova, 
V.A.~Kuzkin, 
Yu.A.~Mochalova
for discussions.}


\appendix
\section{Transport of the probability density}
\label{App-A}
Let wave functions $\psi_\pm=\psi_\mp^\ast$ satisfy complex conjugate Schrödinger-type equations 
\eqref{Schr-eq-pm} and complex conjugate initial conditions \eqref{conjugate-conditions}, 
see Remark~\ref{remark-conj}.
Consider now the transport of the quantity known in the 
framework of the Copenhagen interpretation of  quantum mechanics 
as the quantum probability density:
\begin{equation}
 p\{\psi_+\}=p\{\psi_-\}
\=\lambda{\psi_\pm\psi_\pm^\ast}
=\lambda|\psi_\pm|^2=\lambda{\psi_\pm\psi_\mp}.
\label{E-S-earlier}
\end{equation}
Here
\begin{equation}
  \lambda=
  \left(\iint_{-\infty}^{+\infty}|\psi_\pm(t,\bbr)|^2\,\d S \right)^{-1}
  \label{lambda-def}
\end{equation}
is the normalizing multiplier that is finite if condition \eqref{bcq} is fulfilled.
In  essence, the results of Appendix~\ref{App-A} were obtained by
Schrödinger himself in \cite{Schroedinger1926-AdP4}.

At first, we formulate 
the balance equation for
\begin{equation}
 p_0=\frac p\lambda.
 \label{p0-def}
\end{equation}
Multiplying 
both equations of set \eqref{Schr-eq-pm}
by  $\psi_\mp$ and subtracting the second one from the
first results in
\begin{equation}
  \dot p_0=\frac {\I b}{a}\big(\psi_\pm^\ast\Delta \psi_\pm-\psi_\pm\Delta\psi_\pm^\ast\big)
\end{equation}
or
\begin{gather}
\dot{p_0}=-\bnabla\cdot \bs q_0, 
\label{rho0-balance-earlier}
\\
  \bs q_0
  =\frac {2 b}{a}\Im\big(\psi_\pm^\ast\bnabla\psi_\pm\big).
  \label{q0-def}
\end{gather}

In the quantum framework, it is generally accepted \cite{Schroedinger1926-AdP4,Messiah1-1961} that 
$\bs q\to \bs 0$ as ${\bs r}\to\infty$. Thus, 
the normalizing multiplier $\lambda$ is a constant according to Eq.~\eqref{rho0-balance-earlier}
due to the Ostrogradsky--Gauss theorem. This is a well-known result of 
quantum mechanics. Thus, the equation for balance of the
probability density is
\begin{gather}
\dot{p}=-\bnabla\cdot \bs q, 
\label{rho-balance-earlier}
\\
  \bs q=\bs q_0\lambda,
  \label{q-def}
\end{gather}
where $\bs q$ is the probability current.

\bibliographystyle{plainnat}
\bibliography{bib/all,bib/serge-gost}

\end{document}